\documentclass[eprint,twocolumn,aps,pra,longbibliography,amsmath,amssymb]{revtex4-1}
\usepackage[latin9]{inputenc}
\usepackage{mathrsfs}
\usepackage{amstext}
\usepackage{amssymb}
\usepackage{graphicx}
\usepackage{esint}
\usepackage[export]{adjustbox}[2011/08/13]
\usepackage{hyperref}
\usepackage{pgfplots}
\pgfplotsset{compat=newest}
\usepgfplotslibrary{groupplots}
\usepgfplotslibrary{dateplot}

\makeatletter
\usepackage{epsfig}
\usepackage{bm}
\usepackage{dcolumn}
\usepackage{color}

\makeatother

\begin{document}

\title{Pseudogap regime of a strongly interacting two-dimensional Fermi gas with and without confinement-induced effect range of interactions}

\author{ Brendan C. Mulkerin$^{1}$, Xia-Ji Liu$^{1}$ and Hui Hu$^{1}$} 

\affiliation{$^{1}$Centre for Quantum and Optical Science, Swinburne University
of Technology, Melbourne, Victoria 3122, Australia}

\date{\today}
\begin{abstract}
We investigate theoretically the many-body pairing of a strongly correlated two-dimensional Fermi gas with and without negative confinement-induced effective range. Using a strong-coupling effective field theory in the normal state, we show that the specific heat at constant volume can be used as a characteristic indicator of the crossover from the normal Fermi liquid to the pseudogap state in two dimensions. We calculate the pseudogap formation temperature through the specific heat at constant volume, examining the role of a negative confinement-induced effective range on many-body pairing above the superfluid transition. We compare our results with and without effective range to the recent experimental measurement performed with radio-frequency spectroscopy in Murthy {\it et al.} [Science {\bf 359}, 452-455 (2018)]. Although a good qualitative agreement is found, we are not able to discriminate the effect of the confinement-induced effect range in the experimental data.

\end{abstract}

\maketitle

\section{Introduction}

The role of many-body pairing in Fermi systems above the critical superfluid temperature - the so-called pseudogap pairing - is a complex and intriguing problem. It has been long recognized that the pseudogap pairing is important in underpinning superconductivity in high-temperature superconductors  \cite{loktev2001,Chen2005,Stajic2017}, however due to quantum fluctuations such pairing is difficult to understand \cite{Chien2010,Mueller2017}. The advancement of experimental techniques in trapping and control of interactions in ultracold Fermi gases makes them an ideal platform to study high-temperature many-body pairing across the crossover from a Bose-Einstein condensate (BEC) to a Bardeen-Cooper-Schrieffer (BCS) superfluid \cite{Gaebler2010}. Two-dimensional (2D) ultracold Fermi gases are of particular interest due to the increasingly important role of quantum fluctuations in low dimensions and it is expected that the interaction and temperature regimes where pseudogap pairing dominates, known as the pseudogap regime, is much more pronounced \cite{torma2016,Mueller2017}. 

Probing the pseudogap regime is difficult as there is no conclusive phase transition across the BEC-BCS crossover. The most widely used theoretical definitions of the pseudogap formation temperature are when a minimum enters the density of states (DoS) or there is a "backbend" in the spectral function \cite{Zwerger2009,Ohashi2009,Hu2010prl,Chien2010,Perali2011,Mueller2017}. However, there is no uniquely defined transition and these methods can lead to competing formation temperatures. For example, in 2D the suppression entering the DoS near the Fermi surface leads to a limit in the weakly interacting BCS regime where pairing and condensation occur at different temperatures. Hence, one has to take a more significant suppression in the DoS to define a consistent and meaningful pseudogap formation temperature \cite{Bauer2014}. Another technique to observe the effects  of many-body pairing is to calculate the equation of state (EoS) and thermodynamic properties \cite{kinast2005heat}. It has been observed that the spin susceptibility and specific heat at constant volume contain information about the pairing in a three-dimensional interacting Fermi gas and a characteristic transition temperature can be defined \cite{vanwyk2016}. In this work we will determine the pseudogap regime using the specific heat at constant volume in two dimensions and compare to the pseudogap regime predicted from the suppression in the DoS (see, i.e., Ref.~\cite{Bauer2014}). 

On the experimental side, the advancement of trapping techniques over the last few years has seen a set of important measurements on 2D interacting Fermi gases~\cite{Martiyanov2010,Feld2011,Frolich2011,Zhang2012}. There was much debate about the pairing regime found in these experiments \cite{Watanabe2013,Bauer2014,Marsiglio2015}, where it was argued that the regime probed was not many-body pairing, but two-body pairing. In order for many-body pairing to exist the Fermi gas must have a defined Fermi surface and pairing comes from the many-body nature of the system, which is seen to be true for a chemical potential $\mu>0$. However, it has also been argued by Ref.~\cite{Marsiglio2015} that the criterion of a positive chemical potential is too strict and that many-body pairing can exist for a wider interaction and temperature range. Recent experimental work by Murthy in Ref.~\cite{Murthy2018}  has seemingly observed the high-temperature pairing in 2D Fermi gases for a wide range of interaction strengths and temperatures, although they did not determine a phase diagram.

All previous theoretical methods used to study the pseudogap regime in two dimensions (i.e., $x-y$ plane) have relied upon a \emph{single} channel model of fermions with a contact interaction. It has been found that this model works very well in explaining experimental data for both below the Berezinskii-Kosterlitz-Thouless (BKT) transition \cite{Bertaina2011,He2015,Schonenberg2016,Mulkerin2017,he2019reaching} and the EoS in the normal state \cite{Klimin2012,Watanabe2013,Bauer2014,Mulkerin2015,Marsiglio2015,Matsumoto2014,Anderson2015}. However,  recent measurements on the breathing mode and quantum anomaly of 2D Fermi gases \cite{Holten2018,Peppler2018} have found a significant deviation from the state-of-the-art theoretical prediction using the single-channel model \cite{Hofmann2012,Gao2012}. This difference could not be explained through a temperature dependence of the experimental data alone \cite{Mulkerin2018}, and including higher-order excitations along the $z$-axis of the quasi-2D system is crucial in capturing the reduced breathing mode anomaly \cite{Toniolo2018,Hu2019}. Theoretical studies focused on the importance of the quasi-2D nature of scattering and the increased role of confined fermions being able to occupy  higher excited single-particle states along the $z$-direction, even when the trapping is extremely tight~\cite{Kestner2007b,Fischer2013,Hu2019,Wu2019}. Including dressed molecules within a two-channel model has been found to effectively describe this situation \cite{Levinsen2013,Hu2019}, where the molecular state encapsulates the higher excited states and characterizes a confinement-induced effective range of interactions. This highlights the importance of understanding the pseudogap regime by using the two-channel model. 

The purpose of this work is to understand the role played by the confinement-induced effective range on many-body pairing within the two-channel model. Using a field theoretic method to include pairing fluctuations, we calculate the pseudogap formation temperature from the  specific heat at constant volume, $\tilde{T}$ \cite{vanwyk2016}. We compare this characteristic temperature to the pseudogap temperature determined through a suppression in the DoS at the Fermi surface. We find that when using the definition of the pseudogap temperature, $T^*$, as a dip in the DoS of 25\% of the value at the left fringe, there is a good agreement between $T^{*}$ and $\tilde{T}$ in the weakly interacting regime. We then investigate the role played by the confinement-induced effective range on the pseudogap formation temperature of $\tilde{T}$ and see that the effective range shifts the pseudogap window towards weaker binding energies. Finally, we compare our results to the recent radio-frequency (rf) spectroscopy measurements of the pseudogap regime by Murthy \textit{et al.} in Ref.~\cite{Murthy2018}, which is the most promising way to experimentally map out the pseudogap regime. For this purpose, we also calculate rf-spectra for a trapped system from the analytically continued Green's function, and examine the role of the confinement-induced effective range. A good \emph{qualitative} agreement is found between the experimental data and the theoretical pseudogap regime defined by $\tilde{T}$. However, we find that the inclusion of the confinement-induced effective range does not improve the agreement.

The rest of our manuscript is set out as follows. In Sec.~\ref{sec:hamil}, we introduce the two-channel model Hamiltonian and outline the many-body $T$-matrix theory. In Sec.~\ref{sec:results}, we calculate the specific heat at constant volume for a 2D interacting Fermi gas, and using the properties of the specific heat we determine the pseudogap regime and compare it to the pseudogap regime found from the DoS. In Sec.~\ref{comp}, we compare our results to the recent experimental measurements, by calculating the rf-spectra and the pseudogap temperature. And finally in Sec.~\ref{sec:conc}, we summarize our findings. For simplicity we set $\hbar=1$ throughout.

\section{Hamiltonian} \label{sec:hamil}

We start our calculation of the many-body Green's function within a two-channel model of the 2D interacting Fermi gas in the normal state, described by the Hamiltonian~\cite{Ohashi2002,Gurarie2007,Tajima2018b,Mulkerin2020}:
\begin{eqnarray}\label{eq:hamil}
\mathcal{H} & = & \sum_{\mathbf{k}\sigma}\xi_{\mathbf{k}}c_{\mathbf{k}\sigma}^{\dagger}c_{\mathbf{k}\sigma}^{\phantom{\dagger}}+\sum_{\mathbf{q}}\left(\epsilon_{\mathbf{q}}/2+\nu-2\mu\right)b_{\mathbf{q}}^{\dagger}b_{\mathbf{q}}^{\phantom{\dagger}}\nonumber \\
 &  & +g_{\rm b}\sum_{\mathbf{kq}}\left(b_{\mathbf{q}}^{\phantom{\dagger}}c_{\mathbf{q}/2+\mathbf{k}\uparrow}^{\dagger}c_{\mathbf{q}/2-\mathbf{k}\downarrow}^{\dagger}+ {\rm H.c.}\right),
\end{eqnarray}
where ${\rm H.c.}$ is the Hermitian conjugate, $c_{\mathbf{k}\sigma}$ are the annihilation operators of atoms with spin $\sigma=\uparrow,\downarrow$ and mass $M$ in the open channel, and  $b_{\mathbf{q}}$ are the annihilation operators of molecules in the closed channel.  The kinetic energy of the Fermi atoms measured from the chemical potential $\mu$ is $\xi_{\mathbf{k}} = \epsilon_{\mathbf{k}}-\mu$, where $\epsilon_{\mathbf{k}}=k^2/(2M)$. The threshold energy of the diatomic molecule is $\nu$ and the Feshbach coupling is $g_{\rm b}$. As we have used a momentum-independent Feshbach coupling constant, which is unphysical at the high energy, there is an ultraviolet divergence. This divergence can be removed by renormalizing $\nu$, as we discuss in detail in Appendix A. $\nu$ and $g_{\rm b}$ are related to the physical observables of the binding energy $\varepsilon_B$ and the effective range of interactions $R_s<0$ via,
\begin{alignat}{1}
\nu&=-\varepsilon_B+g^{2}_{\rm b}\sum_{\mathbf{k}}\frac{1}{2\epsilon_{\mathbf{k}}+\varepsilon_B},\\
g^{2}_{\rm b}&=-\frac{4\pi\hbar^{4}}{M^{2}}\frac{1}{R_{s}}.
\end{alignat}

\subsection{Many-body $T$-matrix theory} \label{sec:Tmatrix}

\begin{figure}\centering{}
\includegraphics[width=0.9\columnwidth]{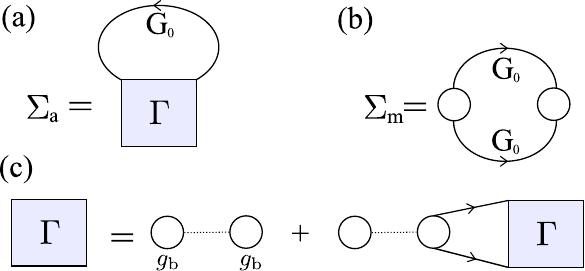}
\caption{\label{fig:diags} (color online) The Feynman diagrams for (a) the fermion self energy, (b) the molecular self-energy, and (c) the vertex function within the ladder-approximation. }
\end{figure}

We consider the effect of pair fluctuations on the normal state properties of a strongly-correlated Fermi system through the non-self-consistent $T$-matrix approximation. The interacting thermal Green's function of fermions at temperature $T$ is given by \cite{Liu2005PRA,Ohashi2009,Tajima2018b}, 
\begin{alignat}{1} \label{eq:gfdown}
G(\mathbf{k},i\omega_m)=\frac{1}{i\omega_m-\left(\epsilon_{\mathbf{k}}-\mu\right)-\Sigma_{a}(\mathbf{k},i\omega_m)},
\end{alignat}
where we sum all of the ladder-type diagrams to obtain the self-energy (see Fig.~\ref{fig:diags}(a)), 
\begin{equation}
\Sigma_{a}=k_{B}T\sum_{\mathbf{q},i\nu_{n}}G^{(0)}\left(\mathbf{q}-\mathbf{k},i\nu_{n}-i\omega_{m}\right)\Gamma\left(\mathbf{q},i\nu_{n}\right).\label{eq:selfenergy1}
\end{equation}
Here, the fermionic and bosonic Matsubara frequencies are respectively $\omega_{m}=(2m+1)\pi k_BT$ and $\nu_{n}\equiv2n\pi k_BT$ for integers $m$ and $n$, and 
the free Green's function is $G^{(0)}(\mathbf{k},i\omega_m)=(i\omega_m-\xi_{\mathbf{k}})^{-1}$. The vertex function $\Gamma\left(\mathbf{q},i\nu_{n}\right)$, which is an effective bosonic propagator, can be written through Fig.~\ref{fig:diags}(c), 
\begin{equation}
\Gamma^{-1}\left(\mathbf{q},i\nu_{n}\right)=U_{\rm eff}^{-1}\left(\mathbf{q},i\nu_{n}\right)+\Pi\left(\mathbf{q},i\nu_{n}\right),\label{eq:vertexfunction}
\end{equation}
with the effective interaction $U_{\rm eff}\equiv g_{\rm b}^2 D_0(\mathbf{q},i\nu_n)$ and the pair propagator $\Pi(\mathbf{q},i\nu_{n})$, 
\begin{alignat}{1}
\Pi &=k_{B}T\sum_{\mathbf{k},i\omega_{m}}G^{(0)}\left(\mathbf{q}-\mathbf{k},i\nu_{n}-i\omega_{m}\right)G^{(0)}\left(\mathbf{k},i\omega_{m}\right), \nonumber \\
&=k_{B}T\sum_{\mathbf{k},i\omega_{m}}\frac{1-f(\xi_{\frac{\mathbf{q}}{2}-\mathbf{k}})-f(\xi_{\frac{\mathbf{q}}{2}+\mathbf{k}})}{2\epsilon_{\mathbf{k}}-2\mu+ \epsilon_{\mathbf{q}}/2-i\nu_n }.
\label{Eq:propagator2p}
\end{alignat}
Here, the free Green's function of a molecular boson is $D_0(\mathbf{q},i\nu_n) = 1/\left[i\nu_n - \epsilon^{\rm B}_{\mathbf{q}} \right]$ with dispersion $\epsilon^{\rm B}_{\mathbf{q}} = \epsilon_{\mathbf{q}}/2 -\nu +2\mu$. As shown in Fig.~\ref{fig:diags}(b), similar to the fermionic Green's function, the interacting Green's function of the molecular boson also includes a self-energy correction,
\begin{alignat}{1}
D(\mathbf{q},i\nu_n) = \frac{1}{i\nu_n  - \epsilon_{\mathbf{q}}/2-\nu +2 \mu- \Sigma_{\rm m}(\mathbf{q},i\nu_n)},
\end{alignat}
where $\Sigma_{\rm m}(\mathbf{q},i\nu_n)$ is given by 
\begin{alignat}{1} \label{eq:mol_se}
\Sigma_{\rm m} = -g_{\rm b}^2 \Pi({\mathbf{q}},i\nu_n).
\end{alignat}
At a given temperature, binding energy and effective range we tune the chemical potential to satisfy the particle number equation:
\begin{alignat}{1} 
N &= N_{\rm a} + 2N_{\rm m} \nonumber \\
&=2k_BT\sum_{\mathbf{k},i\omega_m}G(\mathbf{k},i\omega_m)+2k_BT\sum_{\mathbf{q},i\nu_n}D(\mathbf{q},i\nu_n). \label{eq:Dens:eos}
\end{alignat}
To make the equations dimensionless, we define the Fermi units $k_{\textrm{F}}=(2\pi n)^{1/2}$, $\varepsilon_{\rm F}=k_{\textrm{F}}^{2}/(2M)$, and $T_{\textrm{F}}=\varepsilon_{\rm F}/k_{B}$, where $n=N/V=k_{\rm F}^2/2\pi$ is the total density and $V$ is the area (or the volume in 2D). We then converge the chemical potential $\mu/\varepsilon_{\rm F}$ at a given reduced temperature $T/T_{{\rm F}}$, binding energy $\varepsilon_{B}/\varepsilon_{\rm F}$, and effective range $k_{{\rm F}}^{2}R_{s}$.

The closed set of Eqs.~\eqref{eq:gfdown}-\eqref{eq:mol_se}, can be solved directly with a numerical sum over the Matsubara frequencies, as done in Ref.~\cite{Liu2005PRA}; however within this methodology it is  difficult to numerically continue the thermal Green's function to the real axis, which is needed for obtaining the spectral function and the DoS. Alternatively, we can analytically continue the Matsubara frequencies
to the real axis first, allowing us to directly calculate the analytically continued Green's function~\cite{Marsiglio2015,Pietil2012,Veillette2008}. The thermal Green's function then becomes
\begin{alignat}{1}\label{eq:Green_cont}
G(\mathbf{k},\omega^{+})=\frac{1}{\omega^{+}-\left(\epsilon_{\mathbf{k}}-\mu_{\downarrow}\right)-\Sigma_{a}(\mathbf{k},\omega^{+})},
\end{alignat}
where $\omega^{+}\equiv\omega+i0^{+}$. Using contour integration the self-energy function
takes the form \cite{Rohe2001}, 
\begin{alignat}{1}
\Sigma_{a} & (\mathbf{k},\omega^{+})=\nonumber \\
 & \int\frac{d\mathbf{q}}{(2\pi)^{2}}\frac{d\epsilon}{\pi}\biggl[b(\epsilon)G^{(0)}(\mathbf{k}-\mathbf{q},\epsilon-\omega^{+}){\rm Im}\Gamma(\mathbf{q},\epsilon^{+})\nonumber \\
 & -f(\epsilon){\rm Im}G^{(0)}(\mathbf{k},\epsilon^{+})\Gamma(\mathbf{k}+\mathbf{q},\epsilon+\omega^{+})\biggl],
\end{alignat}
where $f(z)=[\exp(\beta z)+1]^{-1}$ and $b(z)=[\exp(\beta z)-1]^{-1}$ are the Fermi and Bose distributions respectively, with $\beta \equiv 1/(k_BT)$. We then find the imaginary part of the analytically continued self-energy,
\begin{alignat}{1}
{\rm Im}\,\left[\Sigma_{a}(\mathbf{k},\omega)\right]= & \int\frac{d\mathbf{q}}{(2\pi)^{2}}\frac{d\epsilon}{2\pi}\left[b(\epsilon)+f(\epsilon-\omega)\right]\nonumber \\
 & \times{\rm Im}\Gamma(\mathbf{q},\epsilon)\,{\rm Im}G^{(0)}(\mathbf{q}-\mathbf{k},\epsilon-\omega),
\end{alignat}
and we calculate the real part of the self-energy from the Kramers-Kronig
relation, 
\begin{alignat}{1} \label{eq:Sig_real}
{\rm Re}\,\left[\Sigma_a(\mathbf{k},\omega)\right]=\frac{1}{\pi}\mathcal{P}\int_{-\infty}^{\infty}d\omega'\frac{{\rm Im}\left[\Sigma_a(\mathbf{k},\omega')\right]}{\omega'-\omega}.
\end{alignat}
The DoS is calculated by analytically continuing the fermionic Green's function and integrating over the momenta
\begin{alignat}{1}
\rho(\omega) & = -\frac{1}{\pi} \sum_{\mathbf{k}} {\rm Im} G(\mathbf{k},i\omega_m\rightarrow\omega+i0^+), \nonumber \\
& \equiv \sum_{\mathbf{k}} A(\mathbf{k},\omega+i0^+).
\end{alignat} 

It is possible to relate the above many-body $T$-matrix theory to the Nozi\`ere-Schmitt Rink \cite{nozieres1985bose,Mulkerin2019b} approach by truncating the self-energy to the first order, i.e.,
\begin{alignat}{1} \label{eq:NSR_Green}
G(\mathbf{k},i\omega_m)  = G_0(\mathbf{k},i\omega_m)+G_0 \Sigma_a(\mathbf{k},i\omega_m) G_0.
\end{alignat}
This is equivalent to writing the thermodynamic potential for a two-channel model
\begin{alignat}{1}
\Omega = \Omega^{(0)}_{\rm F} + \Omega^{(0)}_{\rm B}- \sum_{\mathbf{q},i\nu_n} \ln \left[ 1+g_{\rm b}^2D_0 \Pi(\mathbf{q},i\nu_n)  \right],
\end{alignat}
where $\Omega^{(0)}_{\rm F} = 2\sum_{\mathbf{k}} \ln( e^{-\beta\epsilon_{\mathbf{k}}} +1)$ is the free fermionic thermodynamic potential and $\Omega^{(0)}_{\rm B} = \sum_{\mathbf{q}} \ln(e^{-\beta\epsilon^{\rm B}_{\mathbf{q}}} - 1)$ is the free bosonic thermodynamic potential. Although the pressure equation of state (EoS) and thermodynamic properties can be calculated from the density equation of state in Eq.~\eqref{eq:Dens:eos} via the Gibbs-Duhem relation \cite{Bauer2014,Mulkerin2015},  we use the NSR approach for the calculation of the specific heat at constant volume as this is considerably simpler: it is more feasible to calculate numerically the derivatives with respect to the chemical potential and interaction strength. We expect that there is a small correction to the specific heat at constant volume when the self-energy becomes more significant and the approximation of Eq.~\eqref{eq:NSR_Green} weakens.

Through the thermodynamic potential we can calculate the thermodynamic properties of the system, starting with the pressure EoS $P=\Omega/V$, the energy $E = -TS+\Omega+\mu N$, and the entropy $S = -\left(\partial\Omega/\partial T \right)_{\mu}$ \cite{Mulkerin2020}. The specific heat at constant volume is given by
\begin{alignat}{1}
C_V &= \left(\frac{\partial E}{\partial T}\right)_{V,N}.
\end{alignat}
For the specific heat at constant volume it is simplest to calculate the derivative of the energy with respect to temperature numerically:
\begin{alignat}{1}
C_V = \frac{E\left[\mu(T+\delta),T+\delta\right]-E\left(\mu(T-\delta),T-\delta\right]}{2\delta},
\end{alignat}
and where we set $\delta=0.01T_F$ \cite{vanwyk2016}.

We note that since the superfluid transition temperature predicted by the Thouless criterion is precisely zero in two dimensions~\cite{Hohenberg1966,loktev2001}, we do not consider the finite-temperature transition in this work. It is also important to note the limitations and benefits of the non-self-consistent $T$-matrix scheme. This $T$-matrix scheme is useful as it is possible to analytically continue the Green's function and directly obtain spectral functions: we do not rely on a numerically unsound procedure. The non-self-consistent $T$-matrix approximation is well defined and works well in the high temperature regime where the interaction strength effectively becomes weaker, in the tightly-bound limit where the binding energy $\epsilon_B \gg \varepsilon_{\rm F}$ and molecules are well-formed, or in the weakly-interacting limit where the binding energy is exponentially small. However, when the interactions between performed molecules are strong, such as in the strongly correlated regime and at sufficiently low temperatures, the chemical potential approaches the binding energy and we expect the non-self-consistent $T$-matrix theory to give incorrect results \cite{Matsumoto2014}. In this work we avoid this problem as we focus on the relatively high temperature regime (i.e., at temperatures larger than a characteristic BKT temperature of $\sim0.1T_F$.)

\begin{figure}\centering{}
\includegraphics[width=1.0\columnwidth]{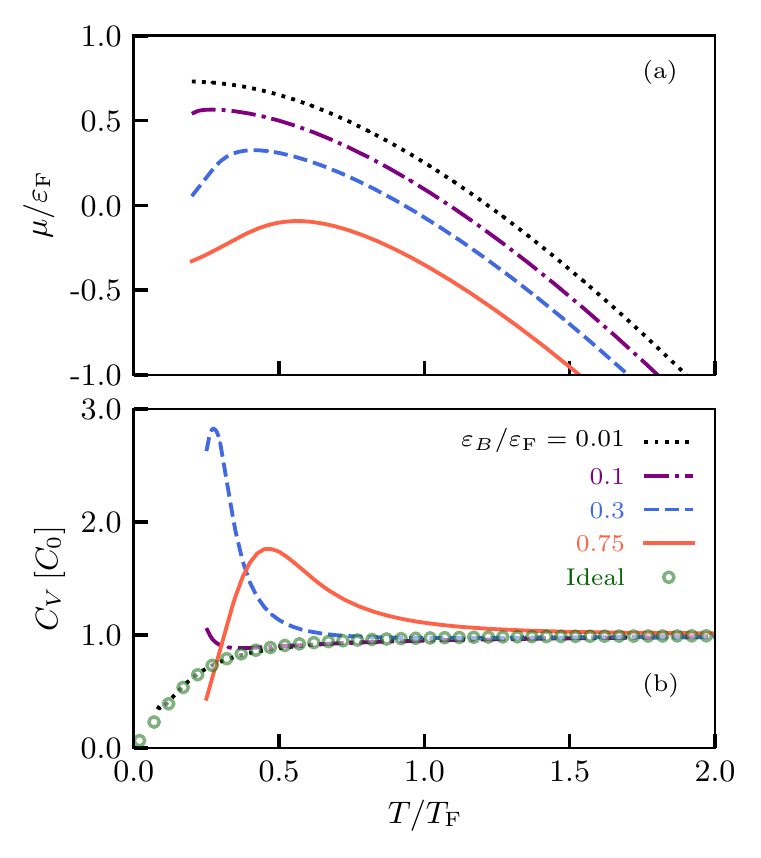}
\caption{\label{fig:Cv_temo} (color online) (a) The chemical potential in units of the Fermi energy for binding energies $\varepsilon_B/\varepsilon_{\rm F}=0.01$ (black dotted), $0.1$ (purple dot-dashed), $0.3$ (blue dashed), and $0.75$ (red solid) and (b) the specific heat at constant volume in units of $C_0=Nk_{\rm B}$ as a function of reduced temperature for the same binding energies. The ideal Fermi gas specific heat predicted by Eq.~(\ref{eq:ideal}) is shown as the symbols.}
\end{figure}

\section{Results} \label{sec:results}

\subsection{specific heat}

 We first consider the broad resonance limit and let $g_{\rm b}\rightarrow\infty$, i.e. $k_{\rm F}^2R_s=0$, in order to understand the general properties of the specific heat at constant volume. Figure~\ref{fig:Cv_temo}(a) shows the reduced chemical potential in units of the Fermi energy as a function of temperature, $T/T_{\rm F}$. We see that the temperature dependence of the chemical potential for each binding energy is non-trivial, and as we go towards the strongly-correlated and low temperature regime we see the chemical potential has a maximum value, indicating the tendency of a transition towards the superfluid state.

In Fig.~\ref{fig:Cv_temo}(b) we plot the specific heat at constant volume in units of $C_0 = Nk_{\rm B} $, as a function of temperature from the weakly-attractive BCS side to the strongly-correlated regime. We see that for the weakest binding energy, $\varepsilon_B/\varepsilon_{\rm F}=0.01$, the specific heat is reduced to the ideal Fermi gas specific heat at constant volume:
\begin{alignat}{1}\label{eq:ideal}
C_V^F = 2\frac{{\rm Li}_{2}\left(-e^{\beta\mu}\right)}{{\rm Li}_{1}\left(-e^{\beta\mu}\right)} - \frac{{\rm Li}_{1}\left(-e^{\beta\mu}\right)}{{\rm Li}_{0}\left(-e^{\beta\mu}\right)}.
\end{alignat}
In the high temperature limit (i.e., $T>T_F$), the specific heat for all interactions is approaching $C_V=Nk_B$. In the relatively high temperature regime (i.e., $T\sim0.3T_F$), the specific heat is enhanced compared to the ideal gas result, and typically exhibits a peak structure. As we move from the weakly-attractive regime to the strongly-coupled regime, the enhancement or peak first increases and then decreases. As we shall discuss in greater detail below, this enhancement connects to the many-body pseudogap pairing. Before doing so, let us briefly review the DoS, which provides a conventional characterization of the pseudogap regime.

\subsection{Density of states}

\begin{figure}\centering{}
\includegraphics[width=1.0\columnwidth]{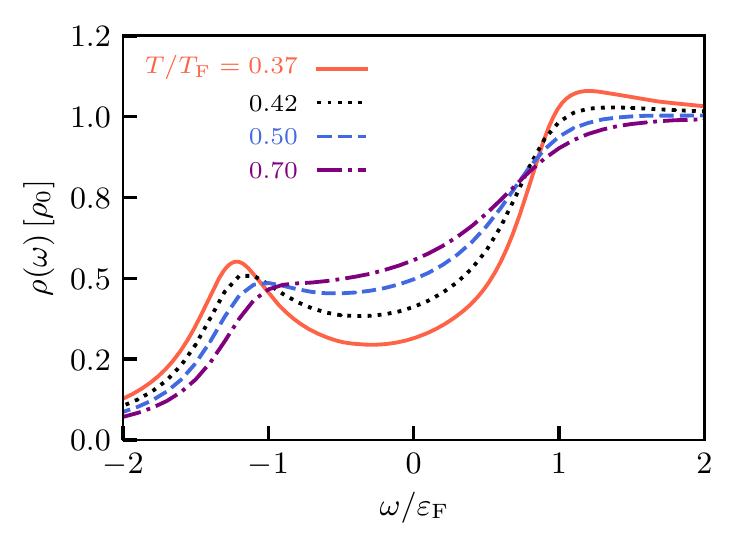}
\caption{\label{fig:Dos} (color online) The density of states is plotted as a function of frequency in units of $\rho_0=m/\pi$  for an interaction strength of $\varepsilon_B/\varepsilon_{\rm F}=0.3$ and a range of temperatures. }
\end{figure}

Indeed, it has been discussed in a range of works that in both two and three dimensions the DoS can be used to find the pseudogap formation temperature \cite{Mueller2017}. In Fig.~\ref{fig:Dos} we plot the DoS at the interaction strength $\varepsilon_B/\varepsilon_{\rm F}=0.3$ for a range of temperatures, normalized by the ideal density of states, $\rho_0=m/\pi$, and showing the evolution of the suppression, or dip, near zero frequency with respect to the chemical potential. It is readily seen that, as the temperature reduces the suppression in the density of states increases. In this work we choose to take the pseudogap formation temperature $T^{*}$ when there is a significant dip near the Fermi surface \cite{Bauer2014}, that is, when the lowest value near $\omega/\varepsilon_{\rm F}\simeq0$ is 25\% lower than the left peak value. In this way, we can approach the BKT transition temperature in the weakly interacting regime, and as the temperature is lowered the system will move directly from a normal Fermi liquid to a superfluid. 

\begin{figure}\centering{}
\includegraphics[width=1.0\columnwidth]{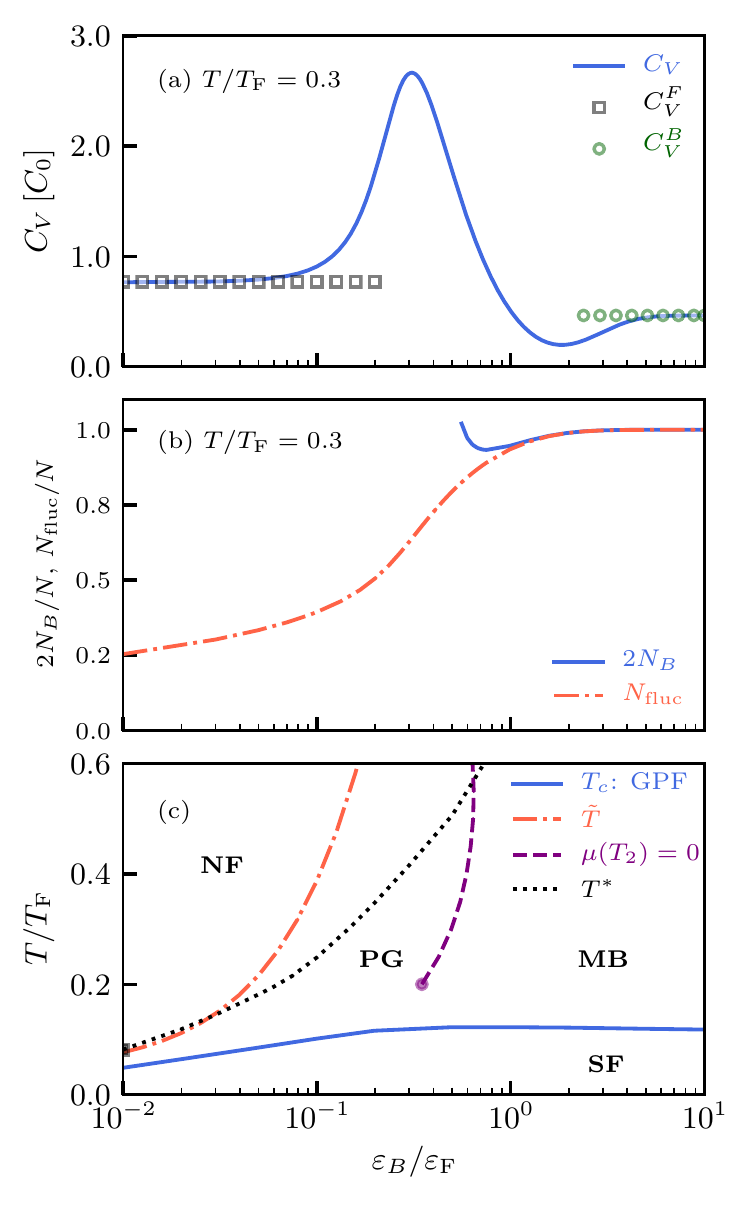}%
\caption{\label{fig:mixed_Cv} (color online) (a) The specific heat at constant volume in units of $C_0=Nk_{\rm B}$ as a function of binding energy $\varepsilon_B/\varepsilon_{\rm F}$. The Fermi and Bose ideal limits are shown square and circular symbols, respectively. (b) The fluctuation contribution to the number equation, $N_{\rm fluc}=-\partial \Omega_{\rm NSR}/\partial \mu$ (red dot-dashed) and twice the number $N_B$ of stable molecules (blue solid). (c) Phase diagram of the 2D Fermi gas as function of binding energy and reduced temperature. Crossover to many-body pairing (PG) from the the normal Fermi gas (NF) found from $C_V$ is given by $\tilde{T}$ (red dot-dashed). $T^*$ (black dashed) is the pseudogap formation temperature found from the density of states. $T_c$ (blue solid)  defines the BKT transition to a superfluid (SF)  and is given by the  Gaussian pair fluctuation theory in Ref.~\cite{Mulkerin2017}. The temperature $T_2$ where $\mu(T_2)=0$ (purple dashed) is the crossover temperature towards a two-body dominated regime.}
\end{figure}

\subsection{Phase diagram}

To understand the enhancement of the specific heat at constant volume in the relatively high temperature regime  we plot in Fig.~\ref{fig:mixed_Cv}(a) $C_V$ as a function of binding energy at a fixed temperature $T/T_{\rm F}=0.3$. We see there is a clear enhancement of $C_V$ peaked at binding energy $\varepsilon_B/\varepsilon_{\rm F}\simeq0.3$, indicating that in this regime there are high-temperature many-body Cooper pairs forming. The specific heat smoothly evolves from an ideal Fermi gas $C_V^F$ on the weakly attractive BCS side to an ideal Bose gas $C_V^B$ of mass $2M$ and density $N/2$  on the strongly attractive BEC side. Here, the ideal Bose gas specific heat at constant volume takes the same form as in Eq.~\eqref{eq:ideal} \cite{Robert1964}; however, the chemical potential is determined using the number equation for an equivalent Bose system with mass $2M$ and density $N/2$. 

To see how these many-body pairs arise, we plot in Fig.~\ref{fig:mixed_Cv}(b) the fluctuation contribution to the total number density, $N_{\rm fluc} = -\partial\Omega_{\rm NSR}/\partial \mu$, as a function of the binding energy (red dot-dashed line), where
\begin{alignat}{1}
\Omega_{\rm NSR} = -\frac{1}{\pi}\sum_{\mathbf{q}} \int_{-\infty}^{\infty} \frac{d\omega}{e^{\beta\omega}-1}\delta(\mathbf{q},\omega),
\end{alignat}
and $\delta(\mathbf{q},\omega)\equiv-{\rm Im}\ln[-\Gamma^{-1}(\mathbf{q},\omega+i0^{+})]$. The contribution of $N_{\rm fluc}$ to the total density can be thought of as renormalized Cooper-pair fluctuation and can be broken into contributions from metastable pairs and scattered states \cite{Ohashi2002,Massignan2008PRA}. In particular, if there is no Fermi surface and the chemical potential is negative, i.e. $\mu<0$, it is possible to divide the fluctuation contribution into twice the number of stable molecules $N_B$ (blue solid line) and of scattered states $N_{\rm sc}$ (not shown in the figure) \cite{Ohashi2002,vanwyk2016}. We plot in Fig.~\ref{fig:mixed_Cv}(b) twice the number of stable molecules $N_B$ for binding energies greater than $\varepsilon_B/\varepsilon_F>0.5$, where $N_B$ can be calculated from the bound state contribution \footnote{See the appendix of Ref.~\cite{vanwyk2016} for retails on the calculation of $N_B$.}. For binding energies below this value the chemical potential is positive and the stable molecule formulation is unphysical. Thus, it is clear that the contribution of pairs below binding energies $\varepsilon_B/\varepsilon_F\sim0.5$ should be from many-body pairing and gives rise to the enhancement of the specific heat at constant volume $C_V$.

Following the idea of Ref.~\cite{vanwyk2016} we take the minimal value of $C_V(T/T_{\rm F})$ as a characteristic transition temperature, $\tilde{T}$, between the normal Fermi gas (NF) and a many-body paired system, i.e. the pseudogap regime (PG). This value signifies the deviation from the ideal $C_V$ in the weakly attractive regime, and breaks down as we approach strongly attractive interactions, and can be seen as the minimum value in Fig.~\ref{fig:Cv_temo}, for temperatures above where the chemical potential is unphysically tending towards the binding energy.
This is not a true transition temperature to the pseudogap regime but a characteristic transition.  

We plot a phase diagram in Fig.~\ref{fig:mixed_Cv}(c) showing the crossover temperature to the pseudogap regime defined by $\tilde{T}$ (red dot-dashed) and $T^*$ (black dotted), and the BKT transition temperature $T_c$ to a superfluid (SF) found by the  Gaussian pair fluctuation theory in Ref.~\cite{Mulkerin2017} (blue solid). We show also the crossover line to a regime dominated by two-body physics by the curve $\mu(T_2)=0$ (purple dashed). This line bounds the pseudogap regime as we increase the binding energy. All together, the three lines of the characteristic temperatures, $\tilde{T}$, $T_c$ and $T_2$, enclose a pseudogap regime. We note that the calculation of $\mu(T_2)=0$ is stopped for temperatures below $T/T_{\rm F}=0.2$ due to the break-down of the NSR and $T$-matrix schemes.

\subsection{Effective range dependence}

\begin{figure}\centering{}
\includegraphics[width=1.0\columnwidth]{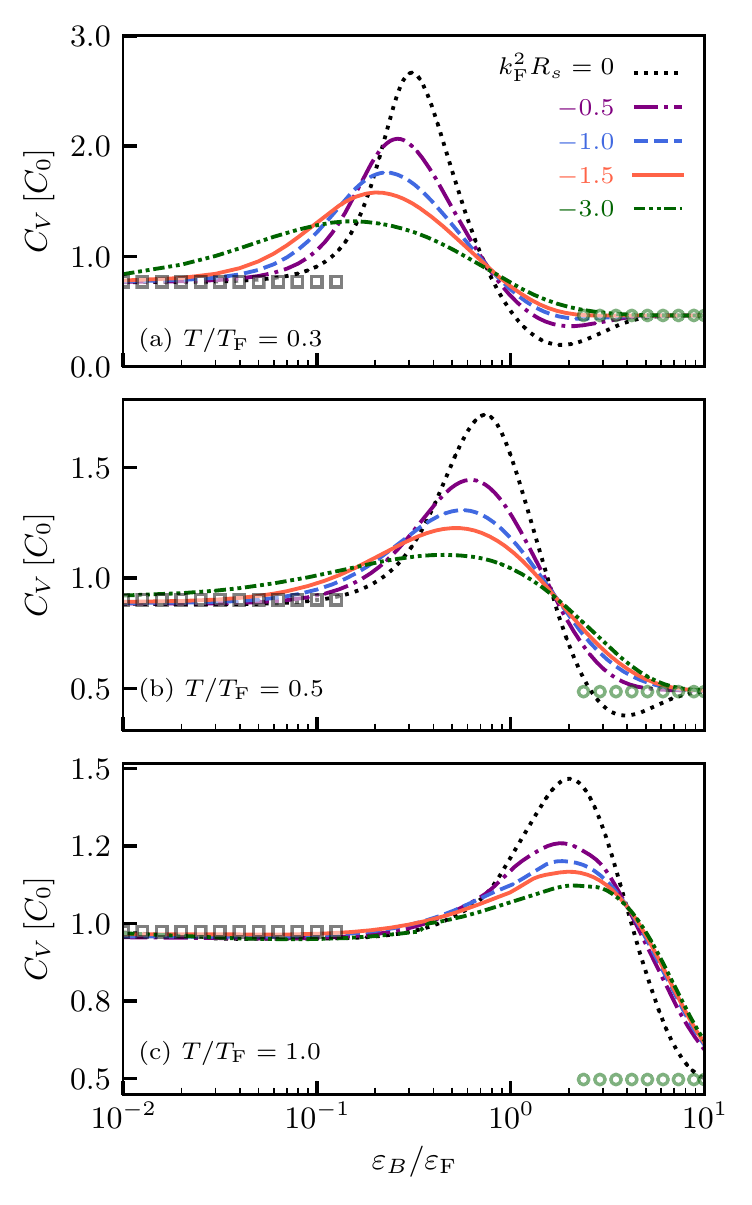}
\caption{\label{fig:Cv_fixedT} (color online) The specific heat at constant volume in units of $C_0=Nk_{\rm B}$ as a function of interaction strength $\varepsilon_b/\varepsilon_{\rm F}$ for negative effective ranges: $k_{\rm F}^2R_s=0$ (black dotted), $k_{\rm F}^2R_s=-0.5$ (purple dot-dashed), $k_{\rm F}^2R_s=-1$ (blue dashed), $k_{\rm F}^2R_s=-1.5$ (red solid), and $k_{\rm F}^2R_s=-3$ (green dot-dot-dashed), for temperatures (a) $T/T_{\rm F}=0.25$, (b) $T/T_{\rm F}=0.5$, (c) $T/T_{\rm F}=1.0$. }
\end{figure}

We now move to consider the confinement-induced effective range dependence of the specific heat at constant volume and of the pseudogap formation temperature. We show $C_V$  in Fig.~\ref{fig:Cv_fixedT} as a function of binding energy, $\varepsilon_B/\varepsilon_{\rm F}$ for a negative effective ranges $k_{\rm F}^2R_s=0$ to $-3$ and temperatures (a) $T/T_{\rm F}=0.25$, (b) $T/T_{\rm F}=0.5$, (c) $T/T_{\rm F}=1.0$. 

The behavior of $C_V$ as a function of decreasing effective range is non-trivial: we find that the enhancement in the middle interaction regime (around $\varepsilon_B/\varepsilon_{\rm F}\simeq0.3$) dampens for each temperature, as the negative effective range decreases. This is most likely due to the system more readily forming bound molecules with decreasing negative effective range. For increasing temperature the peak value is also decreasing, and this is to be expected, as for higher temperatures the role of many-body pairing decreases. We also see that the peak value shifts to larger binding energies at high temperatures as the effective range decreases, due to a non-trivial competition of pair formation with decreasing effective range and high temperatures. Furthermore, in the weakly attractive ($\varepsilon_B/\varepsilon_{\rm F}<0.1$) and tightly bound ($\varepsilon_B/\varepsilon_{\rm F}>5$) limits, the specific heat at constant volume more slowly approaches the ideal gas limits, as the effective range decreases. 

\begin{figure}\centering{}
\includegraphics[width=1.0\columnwidth]{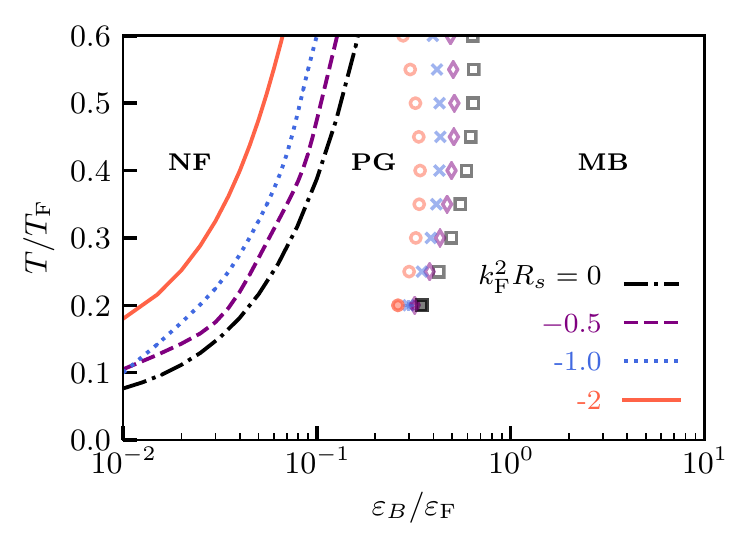}
\caption{\label{fig:maxEb} (color online) The pseudogap formation temperature $\tilde T$ found from the specific heat for a range of negative effective ranges for $k_{\rm F}^2R_s=0$ (black dot-dashed), $k_{\rm F}^2R_s=-0.5$ (purple dashed), $k_{\rm F}^2R_s=-1$ (blue dotted), and $k_{\rm F}^2R_s=-2$ (red solid). We also show the characteristic temperature $T_2$ defined by $\mu(T_2)=0$ using different symbols. At the same effective range, the color is same for lines ($\tilde T$) and symbols ($T_2$).}
\end{figure}

Following the same method to define a pseudogap transition temperature $\tilde{T}$ as in Fig.~\ref{fig:mixed_Cv}, we calculate the effective range dependence of the pseudogap formation and report this main result of our work in Fig.~\ref{fig:maxEb}. The effective ranges are $k^2_{\rm F}R_s=0$ (black dot-dashed), $k^2_{\rm F}R_s=-0.5$ (purple dashed) , $k^2_{\rm F}R_s=-1$ (blue dotted), and $k^2_{\rm F}R_s=-2$ (red solid). We also plot the crossover temperature $T_2$ to a molecule dominated system defined by $\mu(T_2)=0$ using different symbols but the same color for each effective range. The effective range shifts the pseudogap region to weaker binding energies. This is due to the fact that the system more readily forms molecular states with decreasing effective range and increasing binding energy. The interaction window where the pseudogap regime exists remains approximately the same size, however for the smallest effective range ($k_{\rm F}^2R_s=-2$) in the figure, the pseudogap formation temperature is still large for weak interactions. This effect can also be seen in Fig.~\ref{fig:Cv_fixedT}(a), where for decreasing effective range and binding energy, $C_V$ is more slowly approaching the ideal gas result.

\section{Comparison to the experiment} \label{comp}

In this section we outline how to compare our two-channel calculations to the recent experimental observations of Murthy \textit{et al.} in Ref.~\cite{Murthy2018}, with and without the confinement-induced effective range. For this purpose, we include the effect of an inhomogeneous trap through the local density approximation (LDA), $\mu(\mathbf{r}) = \mu_g - \frac{1}{2}M\omega^2 \mathbf{r}^2$, where $\mu_g$ is the global chemical potential, $\omega$ is the trap frequency, and $\mathbf{r}$ is the distance from the center of the trap. We denote the dimensionless radii as $\tilde{r}\equiv r/R_{{\rm TF}}$, $R_{{\rm TF}}^{2}=2k_{{\rm B}}T_{{\rm F}}/(m\omega^{2})$ is the Thomas-Fermi radius for a zero-temperature non-interacting trapped Fermi gas, and the trap Fermi energy $E_{\rm F} = \left(2 N\right)^{1/2}\omega$. 
We find the global chemical potential by enforcing that the total number of atoms satisfies $N=\int d\mathbf{r} n(\mathbf{r})$, where
\begin{alignat}{1}
n(\mathbf{r}) = 2k_BT \int\frac{d\mathbf{k}}{(2\pi)^2} d\omega A(\mathbf{k},\mathbf{r},\omega) n_{\rm F}(\omega),
\end{alignat}
the Fermi distribution is $n_{\rm F}(\omega) = 1/\left(1+e^{-\beta\omega}  \right)$, and $A(\mathbf{k},\mathbf{r},\omega)=(-1/\pi) \, {\rm Im} G(\mathbf{k},\mathbf{r},\omega+i0^+)$ is the spectral function found from the trap dependent Green's function. The inhomogeneous trap means we have trap dependent temperature $T/T_{\rm F}(\mathbf{r})$ and interaction $\ln[k_{\rm F}(\mathbf{r})a_{2D}]$.

The experiment in Ref.~\cite{Murthy2018} measures the local spectral response of a trapped 2D Fermi gas through radio-frequency (rf) spectroscopy. Rf spectroscopy can give information about the properties of the system, by applying a short rf pulse to flip the spin states from an initial strongly interacting system into a weakly-interacting final state and then by measuring the number of transferred atoms. This can then be repeated for a range of detunings of the rf pulse and information about the single-particle properties can be measured. In order to compare the spectra found from the experiment, we calculate the trap dependent rf-spectra. When there is no final state interaction we can take the rf response to be  \cite{Pietil2012,Marsiglio2015}:
\begin{alignat}{1}
I_{\rm rf}(\omega,\mathbf{r})=2\int\frac{d\mathbf{k}}{(2\pi)^{2}}f(\xi_{\mathbf{k},\mathbf{r}}-\omega)A(\mathbf{k},\xi_{\mathbf{k},\mathbf{r}}-\omega),
\end{alignat}
where  $\xi_{\mathbf{k},\mathbf{r}} = \epsilon_{\mathbf{k}}-\mu(\mathbf{r})$. As a self-consistent check to our calculation of the rf spectra, we can calculate the
number density, i.e. 
\begin{alignat}{1}
N=\int_{-\infty}^{\infty} d\omega\int d\mathbf{r}I_{\rm rf}(\omega,\mathbf{r}).
\end{alignat}

To compare our two-channel results to the experimental local rf spectra we need to fix a realistic confinement-induced effective range. This can be done as follows. Using the experimentally measured values of the binding energy and Fermi energy we define the ratio $\varepsilon_B/\varepsilon_{\rm F}$ to obtain the dimensionless effective range for a given interaction.
We require that the two-body $T$-matrix $T_{2B}(E^{+})$ and the quasi-2D scattering
amplitude share the same pole (the same binding energy $\varepsilon_{B}$)
\cite{Wu2019}. It is readily seen 
that the binding energy $\varepsilon_{B}=\kappa^{2}/M$ is related
to the effective range $R_{s}$ by,
\begin{equation}
R_{s}=\frac{2\ln\left(\kappa a_{s}\right)}{\kappa^{2}},\label{eq:Rs}
\end{equation}
where the 2D scattering length $a_s$ is defined in Appendix A. Using the defined binding energy the dimensionless effective range $R_s/a_s^2$ and {\it central} effective range $k_{\rm F}^2R_s$ is then found. 

\begin{figure}\centering{}
\includegraphics[width=1.0\columnwidth]{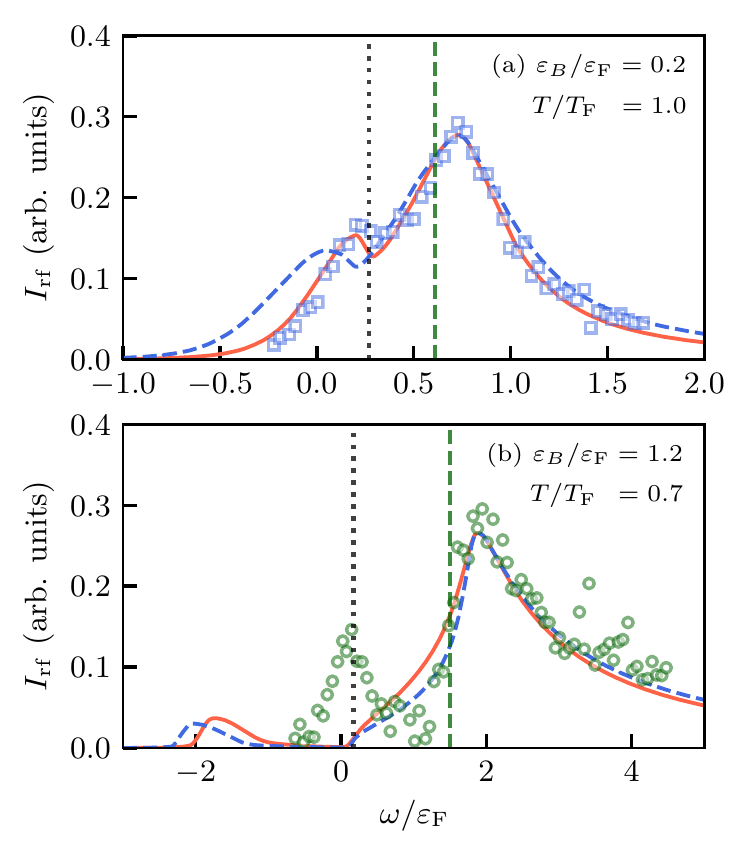}
\caption{\label{fig:spectra} (color online)  Comparison of the local spectra from the $T$-matrix (solid lines), with a finite negative effective range (dashed-line), and experimental results of Ref.~\cite{Murthy2018} (symbols). (a) is for central binding energy $\varepsilon_B/\varepsilon_{\rm F}\simeq0.2$ and local temperature $T/T_{\rm F}=1.0$. (b) is from interaction strength $\varepsilon_B/\varepsilon_{\rm F}\simeq1.21$ and local temperature $T/T_{\rm F}=0.7$. The green dashed lines are the threshold energy and black dotted are the free energy, both determined experimentally.}
\end{figure}

In Fig.~\ref{fig:spectra} we compare the rf spectroscopy found from the $T$-matrix approximation and from Figs. 3(c) and 3(d) of Ref.~\cite{Murthy2018}. We have taken the experimental values of $\varepsilon_B=1.37$kHz in Fig.~\ref{fig:spectra}(a) and $9.31$kHz in Fig.~\ref{fig:spectra}(b), and local Fermi energies $\varepsilon_{\rm F}=6.56$kHz and $7.61$kHz at two fixed radii $\mathbf{r}$ as in the experiment, respectively. This defines a confinement-induced effective range of $R_s/a_{s}^2\simeq-0.2$ and $R_s/a_{s}^2\simeq-1.2$, respectively.These binding energies correspond to Feshbach resonances of 670G and 690G and we use the measured local trap temperatures of $T/T_{\rm F}=1.0$ and $T/T_{\rm F}=0.7$.  Although there is a realistic effective range in the experiment, for comparison in Fig.~\ref{fig:spectra} we also show the theoretical predictions without effective range using red solid lines. 

Quite generally, there are two peaks in the spectra. The right peak, referred to as the pairing peak, comes from the signal of Cooper pairs. The left peak, referred to as the free peak, is contributed from free, unpaired atoms. In order to compare theoretical and experimental spectra, we normalize our spectra to have the same peak value for the pairing peak, and shift the peak to have its maximum at the same frequency.  This shift may minimize the residual final-state effect, which is present in the experiment but is not captured by our theory.

Firstly, the results at smaller binding energy in Fig.~\ref{fig:spectra}(a) match quite well for the whole spectra when there is \emph{no} finite effective range. Using the same fitting method in Ref.~\cite{Murthy2018} to determine the threshold energy, which is the energy required to break a pair, we find that the threshold and free-peak energies are similar to the experimental values. These experimental values are plotted in Fig.~\ref{fig:spectra}(a) using the vertical lines. The ratio of the difference of these energies to the binding energy indicates that we are in the pseudogap regime for this interaction strength and local temperature. When including the finite negative effective range, the agreement between theory and experiment becomes \emph{worse} and the free peak shifts to negative values of the rf frequency. This red shift is due to the chemical potential being slightly lower and the system more easily forming molecular pairs.

For the strongly attractive regime in Fig.~\ref{fig:spectra}(b), we see the spectra match well for the {\it pairing} peak, but not for the {\it free} peak, which is  strongly renormalized by the chemical potential. The threshold energy is then quite similar: there is closer agreement between the theoretical threshold energy with the finite effective range and the experimental threshold energy. If we take the ratio of the difference of the theoretically determined free and threshold energies to the binding energy we would find that for this interaction and temperature we are also in the pseudogap regime, which we would not expect. This is most likely due to the global chemical potential being negative for large interaction strengths, making the free peak shift to negative frequencies \cite{Barth2014}. It is well known that for a large negative chemical potential the Fermi surface is breaking down and two-body bound pairs can form for any binding energy and we are actually not in the pseudogap regime. In this regime the BCS pairing picture gives a fictitious pairing gap as the chemical potential is the gap~\cite{Mueller2017}.  In the experiment in order to measure the free peak they introduce a population imbalance, which creates a broader free peak structure, we do not consider this imbalance in this work, as in the experiment it is only used as a tool to measure the pairing. Experimentally the free peak is then centered around zero rf frequency. 

The comparison of our theoretical results of the rf-spectra with and without effective range to the experimental data suggests that we can hardly follow the experimental procedure to reliably determine the pseudogap regime, by using the theoretically simulated rf-spectra. This is partly due to the fact that, for rf-spectra the many-body $T$-matrix becomes less accurate in the strongly correlated regime where $\varepsilon_B\sim\varepsilon_{\rm F}$. The comparison between theory and experiment is further complicated by the fact that, in the current treatment our theory fails to account for the final-state effect. Thus, at this stage it seems more reliable for us to theoretically determine the pseudogap regime using the specific heat at constant volume.

\begin{figure}\centering{}
\includegraphics[width=1.0\columnwidth]{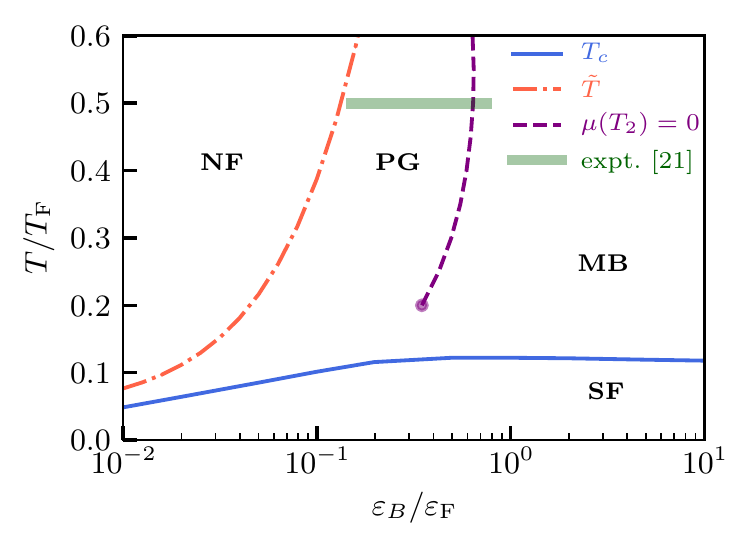}%
\caption{\label{fig:compare_pseudo} (color online) Pseudogap transition temperature phase diagram. The red dot-dashed curve is the specific heat prediction, purple dashed is the curve where the chemical potential becomes negative, the blue solid curve is the BKT transition temperature from the GPF calculation, which are in units of the homogeneous Fermi temperature and energy. }
\end{figure}

In Fig.~\ref{fig:compare_pseudo}, we re-plot the phase diagram for the pseudogap regime found from the specific heat at constant volume at zero effective range and compare it to the experimental result (see, i.e., Fig. 4(b) in Ref.~\cite{Murthy2018}). Here, we do not consider the effective range, since the effect of the effective range does not unambiguously show up  in the rf-spectra as we have just discussed. From the figure, we see that the experimental result at $T\sim0.5T_F$ agrees well with the predicted pseudogap regime. Experimentally the confinement-induced effective range $k_{\rm F}^2R_s$ changes as a function of the binding energy and trap temperature, so it is difficult to have a defined effective range for the entire crossover regime.  We would expect that not considering a finite negative effective range to be reasonable in the weakly interacting regime and as the binding energy increases we would expect the negative confinement-induced effective range to become more important and shift the upper and lower bounds of the pseudogap transition towards smaller binding energies.

\section{Conclusions} \label{sec:conc}

In summary, we have explored the pseudogap regime of a strongly interacting Fermi gas confined to two dimensions with and without a negative confinement-induced effective range. Using the specific heat at constant volume as a probe for high-temperature many-body pairing we have found that in two-dimensions it can be used to determine a good characteristic pseudogap formation temperature when compared to the traditional method of defining the pseudogap regime through a suppression in the density of states. We have seen that, as the effective range decreases, the pseudogap regime shifts to weaker binding energies as the system more preferentially forms pairs. 

By comparing our calculations to the recent experiment of Ref.~\cite{Murthy2018}, we have obtained good qualitative agreement. Plotting directly the measured in-trap radio-frequency spectra, we have shown our results match well the experimental data in the pseudogap regime, and in the strongly-correlated regime the differences can be understood. We have also shown that at high temperatures the many-body pairing regime experimentally defined through radio-frequency measurements fits well with the pseudogap regime theoretically determined from the specific heat at constant volume. However, under the current experimental conditions, it seems difficult to clearly discriminate the effect of the confinement-induced effect range in the radio-frequency spectra and on the pseudogap window, largely due to the insufficient theoretical accuracy for the radio-frequency spectra and insufficient experimental resolution.

\begin{acknowledgments}
Our research was supported by Australian Research Council's (ARC) Discovery Projects:
DP140100637 and FT140100003 (XJL), FT130100815 and DP170104008 (HH).
\end{acknowledgments}

\appendix
\section{Two-body scattering} \label{app:two_body}

Here, we solve the two-particle problem and renormalize the threshold detuning $\nu$. For this purpose, we seek to write the detuning $\nu$ and the channel coupling $g_{\rm b}$ in terms of physical observables, by comparing the two-body $T$-matrix to the quasi-2D scattering amplitude. The two-body $T$-matrix in vacuum is
($E^{+}\equiv k^{2}/M+i0^{+}$),
\begin{equation}
T_{2B}^{-1}\left(E^{+}\right)=U_{eff}^{-1}\left(E^{+}\right)+\sum_{\mathbf{p}}\frac{1}{2\epsilon_{\mathbf{p}}-E^{+}},
\end{equation}
where $\epsilon_{\mathbf{p}}\equiv\hbar^{2}\mathbf{p}^{2}/(2M)$ and
the effective interaction in the presence of the channel coupling
is given by
\begin{equation}
U_{\text{eff}}\left(E^{+}\right)=\frac{g^{2}_{\rm b}}{E^{+}-\nu}.
\end{equation}
Using a large momentum cut-off $\Lambda\rightarrow\infty$ in the integral, we find that
\begin{alignat}{1}
T_{2B}^{-1}\left(E^{+}\right)= \frac{k^2/M-\nu}{g^2_{\rm b}}+\frac{M}{4\pi} \left( \ln \left[\frac{\Lambda^2-k^2}{k^2}\right]+i\pi\right).
\end{alignat}
In the limit of $k\rightarrow0$, we would have $T_{2B}(E^{+})=(\hbar^{2}/M) f_{Q2D}\left(k\right)$. Thus, we consider the low-energy expansion of the quasi-2D scattering amplitude
in \cite{Petrov2001,Bloch2008}:
\begin{alignat}{1}
f_{Q2D}(k\rightarrow 0) = \frac{4\pi}{\sqrt{2\pi}a_{z}/a_{3D}+\varpi\left(k^{2}a_{z}^{2}/2\right)}\label{eq:fQ2D},
\end{alignat}
where $a_z\equiv\sqrt{1/(M\omega_z}$ is the harmonic oscillator length, $a_{3D}$ is the 3D $s$-wave scattering length, and the function $\varpi(x)$ has the form $\varpi(x\rightarrow0) \simeq -\ln(2\pi x/\mathcal{B}) + 2 x\ln 2 +i\pi  $ for $\mathcal{B}=0.9049$. This leads to
\begin{alignat}{1}
T_{2B}\left(E^{+}\right) = \frac{m}{4\pi} \left( -2\ln\left[ka_{s}\right] -R_sk^2+i\pi  \right),
\end{alignat}
where $a_s\equiv a_z\sqrt{\pi/\mathcal{B}}\text{exp}(-\sqrt{\pi/2}a_z/a_{3D})$ is the 2D $s$-wave scattering length and the detuning and Feshbach coupling satisfy
\begin{equation}
a_{s}=\frac{1}{\Lambda}e^{\frac{2\pi\nu}{g_{\rm b}M}},\,\,R_{s}=-\frac{4\pi\hbar^{4}}{M^{2}}\frac{1}{g^{2}_{\rm b}}.
\end{equation}
We remove the cut-off $\Lambda$ by considering the pole
of the two-body $T$-matrix $T_{2B}(E)$, 
$E=E_{B}$, finding that,
\begin{equation}
\nu=E_{B}+g^{2}_{\rm b}\sum_{\mathbf{k}}\frac{1}{2\epsilon_{\mathbf{k}}-E_{B}}.
\end{equation}
The binding energy can be set by $\varepsilon_B=-E_B=\kappa^2/M$ where we have set $k\rightarrow i\kappa$. The effective interaction strength is then
\begin{alignat}{1}\label{eq:Ueff}
\frac{1}{U_{\rm eff}} = - \sum_{\mathbf{k}} \frac{1}{2\epsilon_{\mathbf{k}}+\varepsilon_B} - \frac{M^2 R_s}{4\pi}\left( i\nu_n   -\frac{\epsilon_{\mathbf{q}}}{2}+2\mu+\varepsilon_B \right).
\end{alignat}

\bibliography{two_channel}
\end{document}